# AUTOMATED GENERALISATION OF BUILDINGS USING CartAGen PLATFORM

J. Boodala [a,*], O. Dikshit[a], N. Balasubramanian[a]

[a]Geoinformatics, Department of Civil Engineering
Indian Institute of Technology Kanpur, Kanpur - 208016, Uttar Pradesh, India
jaggubreddy@gmail.com, (onkar, nagaraj)@iitk.ac.in

**ABSTRACT**

In this paper, we present a methodology to automatically derive the generalised representations of buildings at scales 1:25K, 1:50K, and to delineate the urban area for 1:250K scale representation. These generalised representations are derived from 1:10K scale. The automatic generalisation processes are realised using the specific algorithms and the generalisation models available in the CartAGen (CARTographic Agent GENeralisation) platform. The CartAGen is an open source map generalisation platform developed by IGN France. The proposed methodology in this paper is evaluated using the data products available from the Ordnance Survey, UK, and the Survey of India, India. This study investigates the applicability of the CartAGen platform for generalising the data products which have been excluded from the investigations by IGN France. This paper discusses the modifications required for such data products.

## 1. INTRODUCTION

The generalisation is a process of deriving less detailed information from the more detailed ones. The research in generalisation is mainly carried out by the National Mapping Agencies (NMAs) (Duchêne et al., 2014). The automation of generalisation process is of prime importance to NMAs. The automation helps to produce data products of different scales in less time and hence reduces the cost. The NMAs which are actively involved in the automation of generalisation process are Ordnance Survey (UK), IGN France, USGS (USA), ICC (Catalonia), Kadaster (Netherlands), Adv (Germany) and Swisstopo (Switzerland) (Duchêne et al., 2014). The current scope of research in generalisation also finds its application in deriving application dependent generalised products, in developing spatial data infrastructures, etc. These benefits of generalisation make this an important research problem to be solved.

The remaining part of this paper is organised as follows. Section 2 defines the objectives of this research work and the data products used. The methodology followed to achieve the objectives is explained in section 3. The results are delineated and discussed in section 4. Finally, the conclusions are drawn and the scope for future work are discussed in section 5.

## 2. OBJECTIVES AND DATA PRODUCTS USED

The objectives of this study are to derive generalised representations of buildings at scales 1:25K, 1:50K and to delineate the urban area for 1:250K scale representation. Figure 1 shows the structure of the derivation process evaluated in this paper. The target scales, i.e., 1:25K, 1:50K and 1:250K are derived using automatic generalisation from the source data which is of 1:10K scale. Table 1 lists the data products which are used in this study. The OS OpenData products from Ordnance Survey, UK,

---

are used to derive 1:25K, 1:250K representations and also to validate the results. However, the data products from Survey of India are used to derive 1:50K, 1:250K representations. The same data products are also used for validating the results.

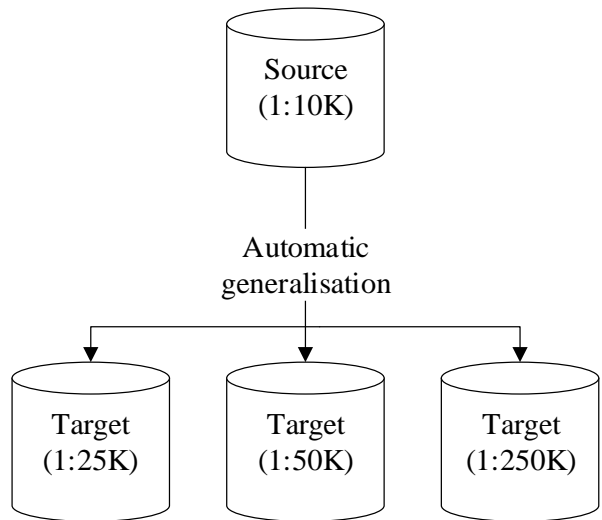

Figure 1. The target scales for automatic generalisation.

Table 1. Data products used.

| Organisation | Product | | | | Purpose |
|---|---|---|---|---|---|
| | Name | Feature type | Scale | Region | |
| Ordnance Survey (OS), UK | OS Open Map - Local (OML) | Building | 1:10K | Oxford district | Source |
| | OS VectorMap District (VMD) | Building | 1:25K | Oxford district | Validation |
| | Strategi | Urban | 1:250K | Oxford district | Validation |
| Survey of India (SoI), India | SoI 10K | Building | 1:10K | Kanpur city | Source |
| | SoI 50K | Block | 1:50K | Kanpur city | Validation |

### 3. METHODOLOGY

To achieve the aforementioned objectives, as shown in Figure 1, the CartAGen (CARTographic Agent GENeralisation) (Renard, Gaffuri, & Duchêne, 2010) platform is used. The CartAGen is an open source map generalisation platform developed by IGN France. This platform includes various generalisation algorithms, multi-agent models and also offers the possibility of building on the existing platform (Touya, Lokhat, & Duchêne, 2019). Figure 2 illustrates the application of the CartAGen platform that is used in this research.

**3.1 Merging Operation**

The merging is defined as a process by which the related features are replaced by a single feature of the same dimensionality, i.e., polygons are merged to form a single polygon (Roth, Brewer, & Stryker, 2011). An algorithm (Damen, van Kreveld, & Spaan, 2008) to merge a set of close buildings in the source data (1:10K) is used to derive 1:25K and 1:50K scale representations. This algorithm is based on morphological operators. The closure followed by opening (i.e., dilation, erosion, erosion and dilation) is applied to derive the generalised representations of buildings. These morphological operations are followed by an edge simplification step to remove the short edges (Damen et al., 2008).

The merging algorithm takes two parameters; (1) buffer size (m) for dilation or erosion operations, (2) edge length (m) for edge simplification.

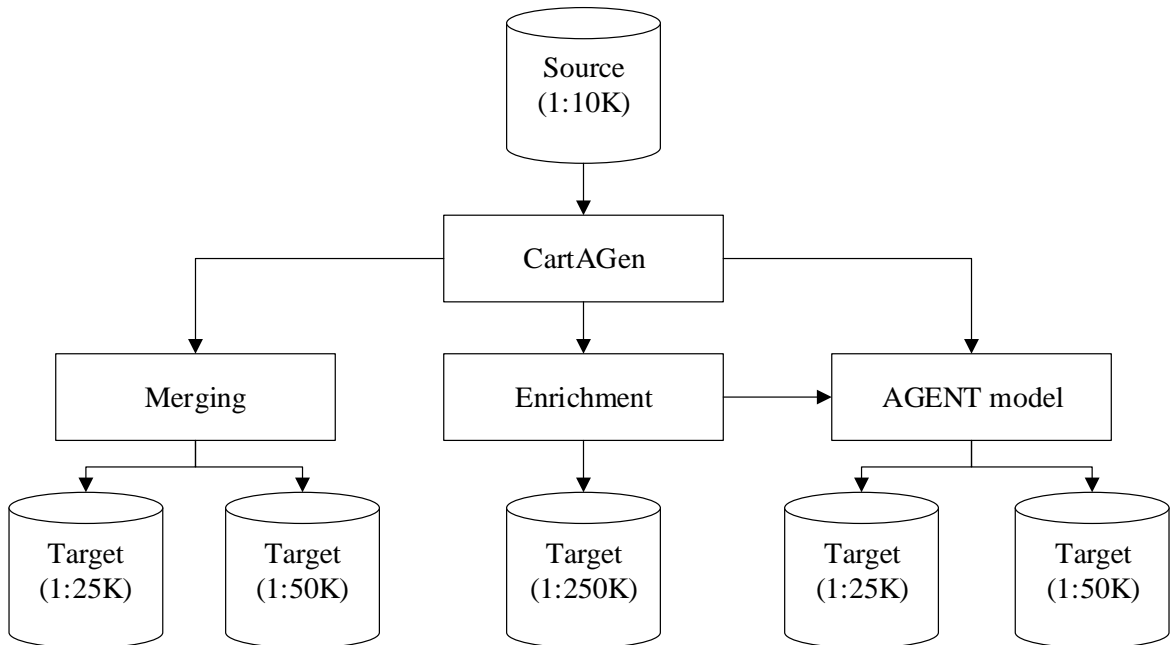

Figure 2. The generalisation process using the CartAGen platform.

## 3.2 Enrichment Operation

The input data (i.e., buildings) is used for the enrichment process to delineate the urban area. This urban area feature can be used for 1:250K scale representation to depict the town/city boundaries. The data enrichment process supports characterisation of the input data, which is necessary for automatic generalisation (Touya et al., 2019), e.g., AGENT model. The urban area delimiting algorithm available in CartAGen platform is proposed by Boffet (2001). It considers the minimum size of the town as a generalisation constraint.

## 3.3 AGENT Model

The CartAGen offers several multi-agents models like AGENT (Ruas, 1999), CartACom (Cartographic generalisation with Communicating Agents) (Duchêne, Ruas, & Cambier, 2012), GAEL (Generalisation based on Agents and Elasticity) (Gaffuri, 2007), etc. However, the AGENT model is suitable for urban areas (Duchêne et al., 2012). Hence, this model is considered for the investigations of the generalisation process with the data products listed in Table 1.

In the AGENT model, each geographic entity (for e.g. building, block) is modelled as an agent. The constraints are defined for each agent. If any of these constraints are not satisfied, then appropriate generalisation action is taken to improve the satisfaction (Ruas & Duchêne, 2007). The agents which take care of their own generalisation are called micro agents (e.g. building) and the agents that control the generalisation of a set of agents are called meso agents (e.g. block agent composed of building agents). Figure 3 shows the hierarchical structure of agents in the AGENT model.

The constraints defined for micro agents are size, granularity, squareness, convexity and elongation. Similarly, those defined for meso agents are block density, big building preservation and proximity. If an agent fails to satisfy the constraint, then the appropriate generalisation algorithm is selected in the AGENT model to improve satisfaction. The input data is enriched to create the block feature, which acts as a meso agent. After the generalisation of buildings, if there are any overlapping buildings that exist then the block (meso agent), which contains these overlapping buildings, will handle the conflict. In the AGENT model, the agents communicate in a hierarchical way, and there is no communication between agents of the same level (Duchêne et al., 2012).

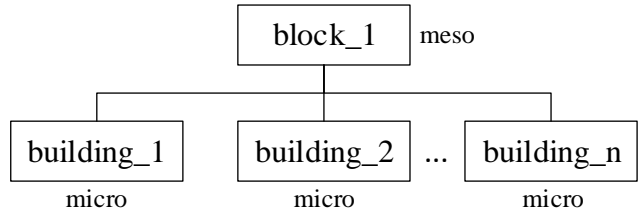

Figure 3. Agents in the AGENT model.

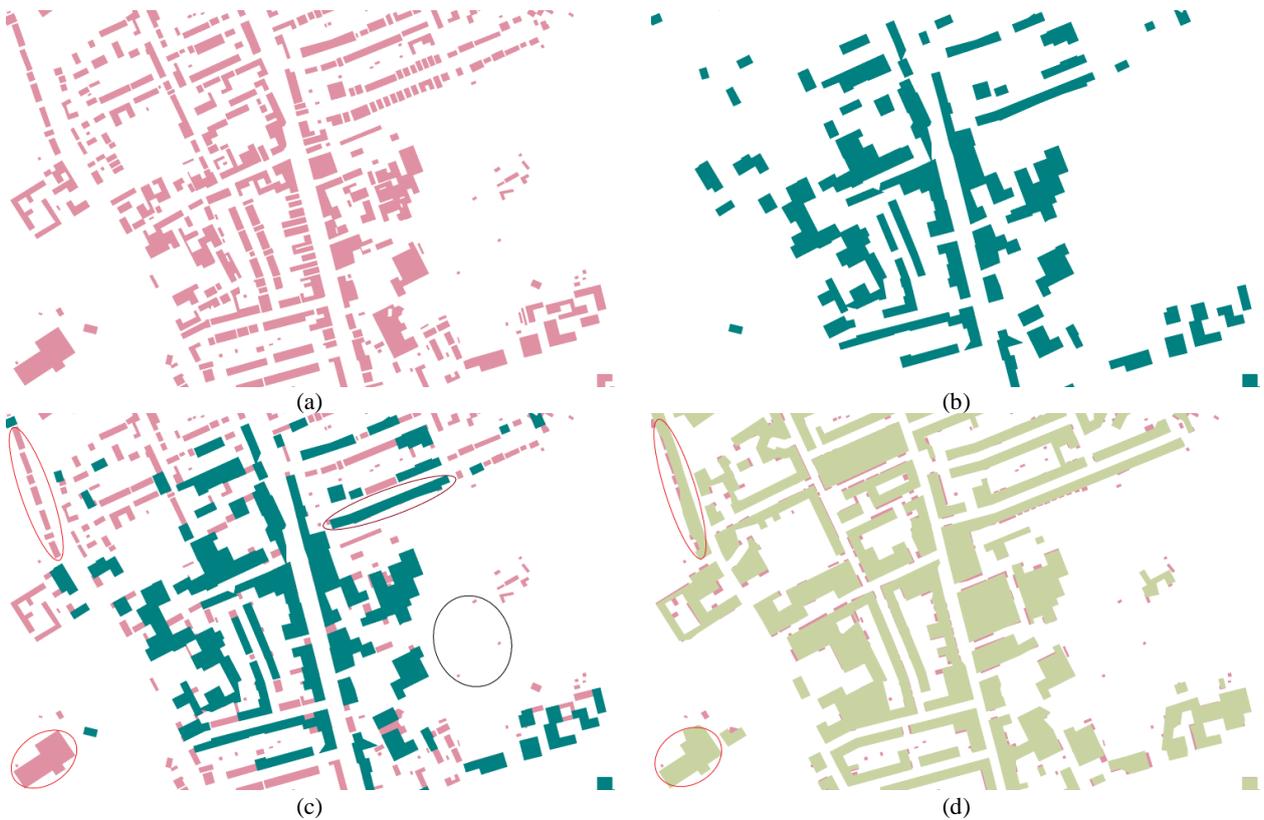

Figure 4. (a) Input OML data, (b) Output of the merging operation, (c) Output of the merging overlaid on OML, (d) VMD overlaid on OML. *Contains OS data © Crown copyright and database right (2017)*.

## 4. RESULTS AND DISCUSSIONS

### 4.1 Results of Merging Operation

Figure 4 shows the results of the merging algorithm using 7 m as the buffer size and 1 m for edge simplification. These parameter values were chosen empirically. The algorithm has removed small isolated buildings (highlighted in the black ring in Figure 4 (c)). It has created a merged building from a set of nearby buildings (highlighted in the brown ring in Figure 4 (c)). The algorithm has failed to retain some of the big individual buildings and also to merge nearby buildings (highlighted in the red ring in Figure 4 (c)). Figure 4 (d) shows the VMD data product which is used for validation. By comparing Figures 4 (c) and (d), it is evident that the algorithm has to be improved for OS data products. Alternatively, a different algorithm has to be tested or developed for this data.

The same merging algorithm is tested on SoI data product. Figure 5 illustrates the results obtained using 6 m as the buffer size and 1 m for edge simplification. These parameter values were chosen empirically. The algorithm is able to detect and drop the small buildings (highlighted in the black ring in Figure 5 (c)). Comparing the results with the SoI 50K data, the merging algorithm is able to retain the shapes of merged buildings corresponding to the source data (highlighted in the green ring in Figures 5 (c) and (d)). Further, the results can be improved by smoothing the boundaries of buildings.

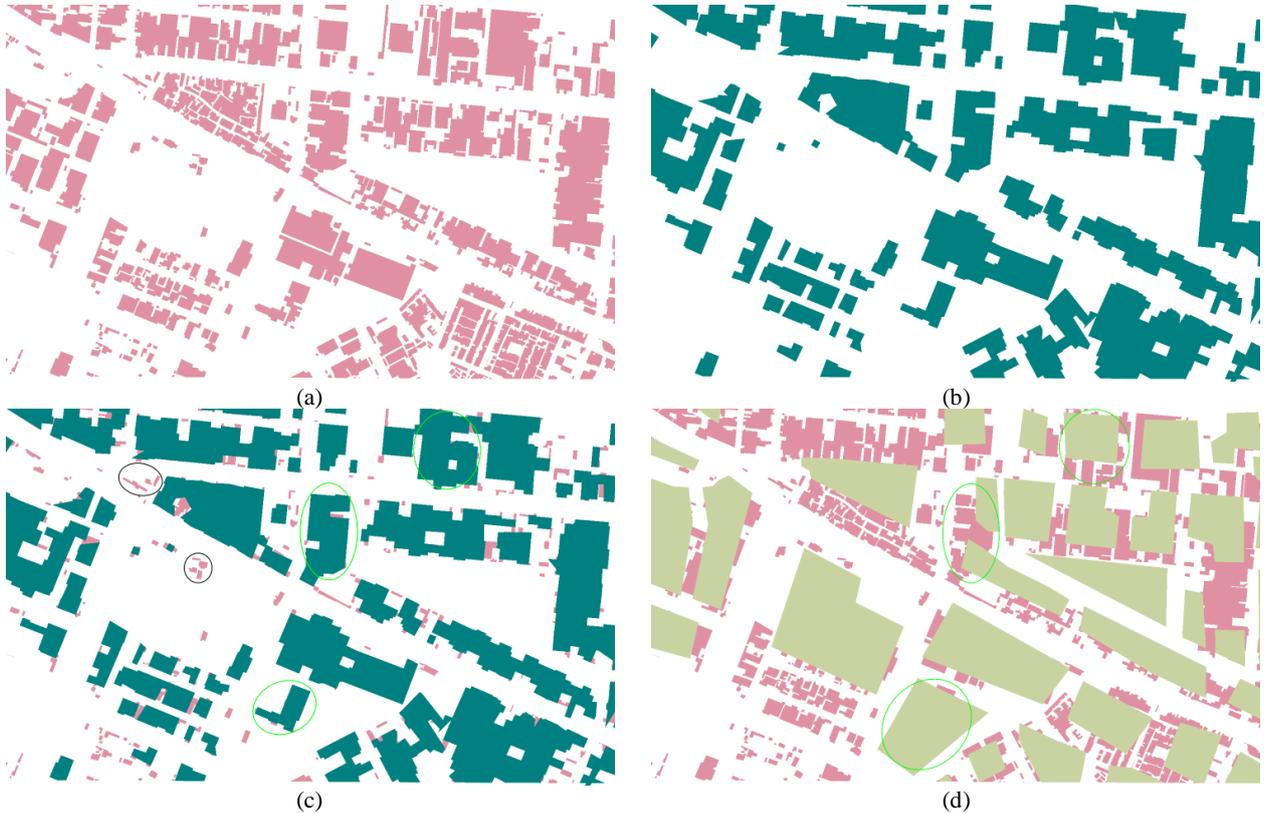

Figure 5. (a) Input SoI 10K data, (b) Output of the merging operation, (c) Output of the merging overlaid on SoI 10K, (d) SoI 50K overlaid on SoI 10K.

## 4.2 Results of Enrichment Operation

Figures 6 and 7 show the results of enrichment of OML and SoI 10K data products, respectively. This enrichment results in the delineation of urban areas (Figures 6 (c) and 7 (b)) using corresponding building features. The algorithm used to delineate the urban areas has successfully dropped the isolated small buildings (highlighted in the black ring in Figures 6 (c) and 7 (b)). The urban areas delimited in Strategi product are more than that of the results of the delimiting algorithm (Boffet, 2001) (highlighted in the red ring in Figures 6 (c) and (d)). This is due to the minimum size of the town constraint used by the delimiting algorithm. In this case, it was set to 750000 m$^2$. It is to be noticed that the boundary of the derived urban area (Figures 6 (c) and 7 (b)) has to be further simplified and smoothed in order to get better visual clarity at 1:250K.

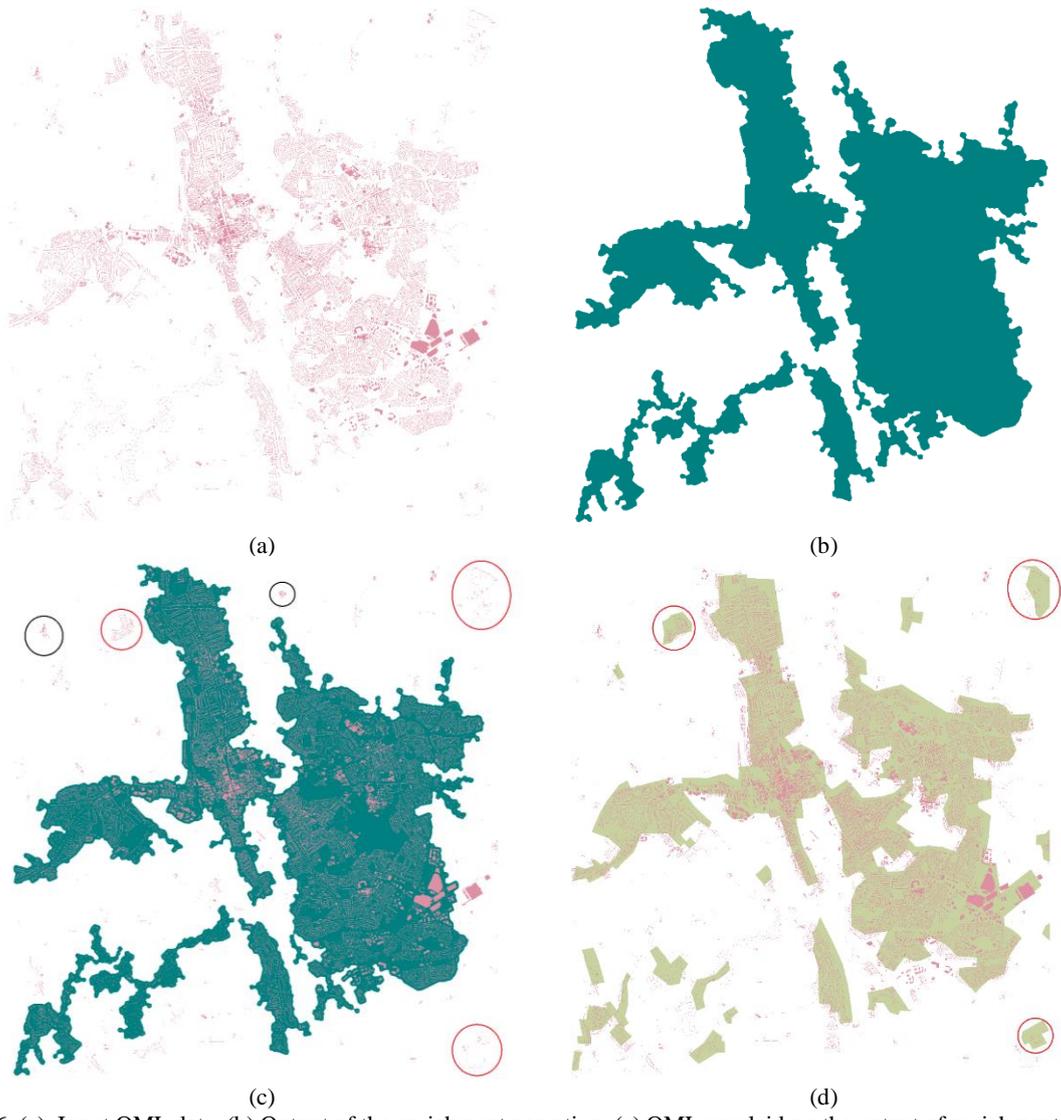

Figure 6. (a) Input OML data, (b) Output of the enrichment operation, (c) OML overlaid on the output of enrichment, (d) OML overlaid on Strategi. *Contains OS data © Crown copyright and database right (2017 & 2015)*.

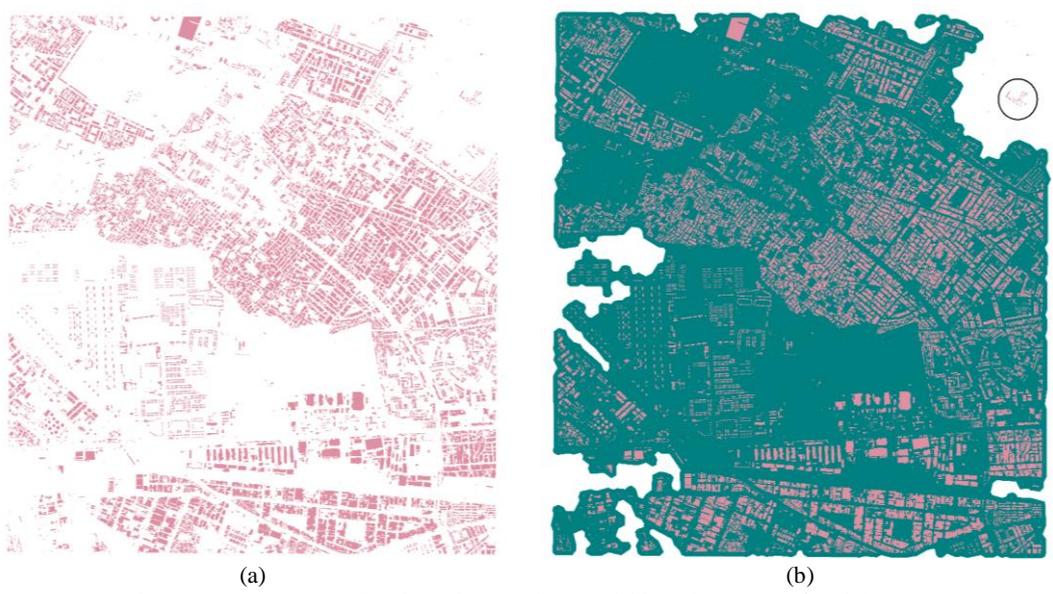

Figure 7. (a) Input SoI 10K data, (b) SoI 10K overlaid on the output of enrichment.

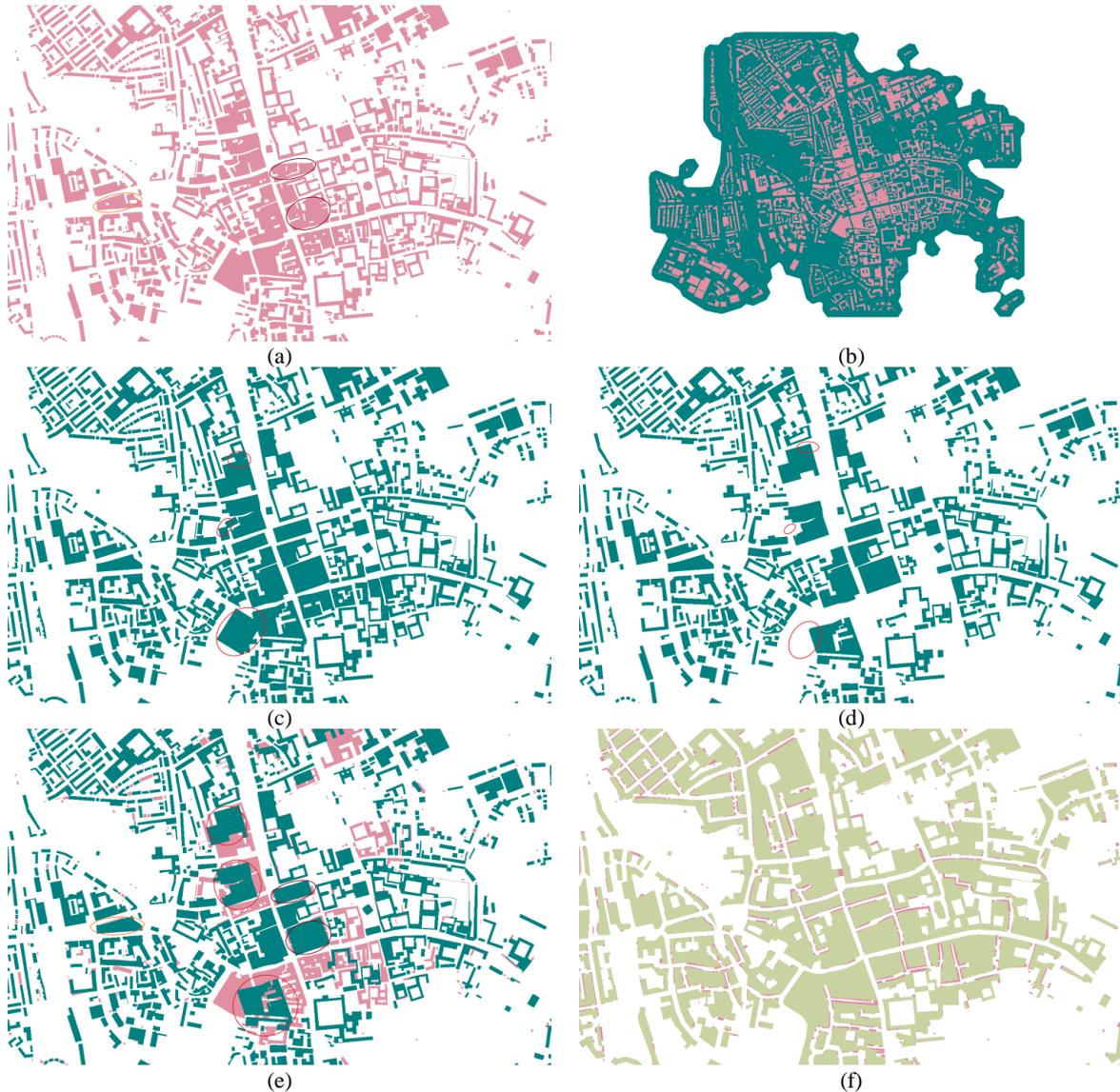

Figure 8. (a) Input OML data, (b) Block created using enrichment, (c) Output of the AGENT model using building micro agents, (d) Output of the AGENT model using block meso agent, (e) Output of the AGENT model overlaid on OML, (f) VMD overlaid on OML. *Contains OS data © Crown copyright and database right (2017).*

## 4.3 Results of AGENT model

Figures 8 and 9 illustrate the results obtained after generalisation using the AGENT model to derive 1:25K and 1:50K representations respectively. In both cases, the block feature (Figures 8 (b) and 9 (b)) is created by an enrichment process and it acts as a meso agent. After the enrichment process, the generalisation is initially carried out on buildings (micro agents). The results obtained after this step are shown in Figures 8 (c) and 9 (c). In this step, each building (micro agent) tries to satisfy its constraints and does not communicate with its neighbouring buildings (micro agents). This may result in the overlapping of the buildings due to generalisation (highlighted in the red ring in Figures 8 (c) and 9 (c)). To solve this overlapping conflict, the block (meso agent) is then generalised. The overlapping buildings within the block are handled by this meso agent. The results after generalising the block are shown in Figures 8 (d) and 9 (d), and resolved conflicts are highlighted in the red ring in both figures. Figures 8 (e) and 9 (e) show the results of generalisation using the AGENT model where newly created and displaced features are seen after resolving the overlapping conflicts (highlighted in the red ring in Figures 8 (e) and 9 (e)). The squared and granularity reduced buildings are highlighted in brown and orange rings in Figures 8 (e) and 9 (e), which can be compared with the input data highlighted using the respective colour rings in Figures 8 (a) and 9 (a). The results of the AGENT model (Figures 8 (e) and 9 (e)) compared to that of OS and SoI data products (Figures 8(f) and 9 (f)) need to be improved.

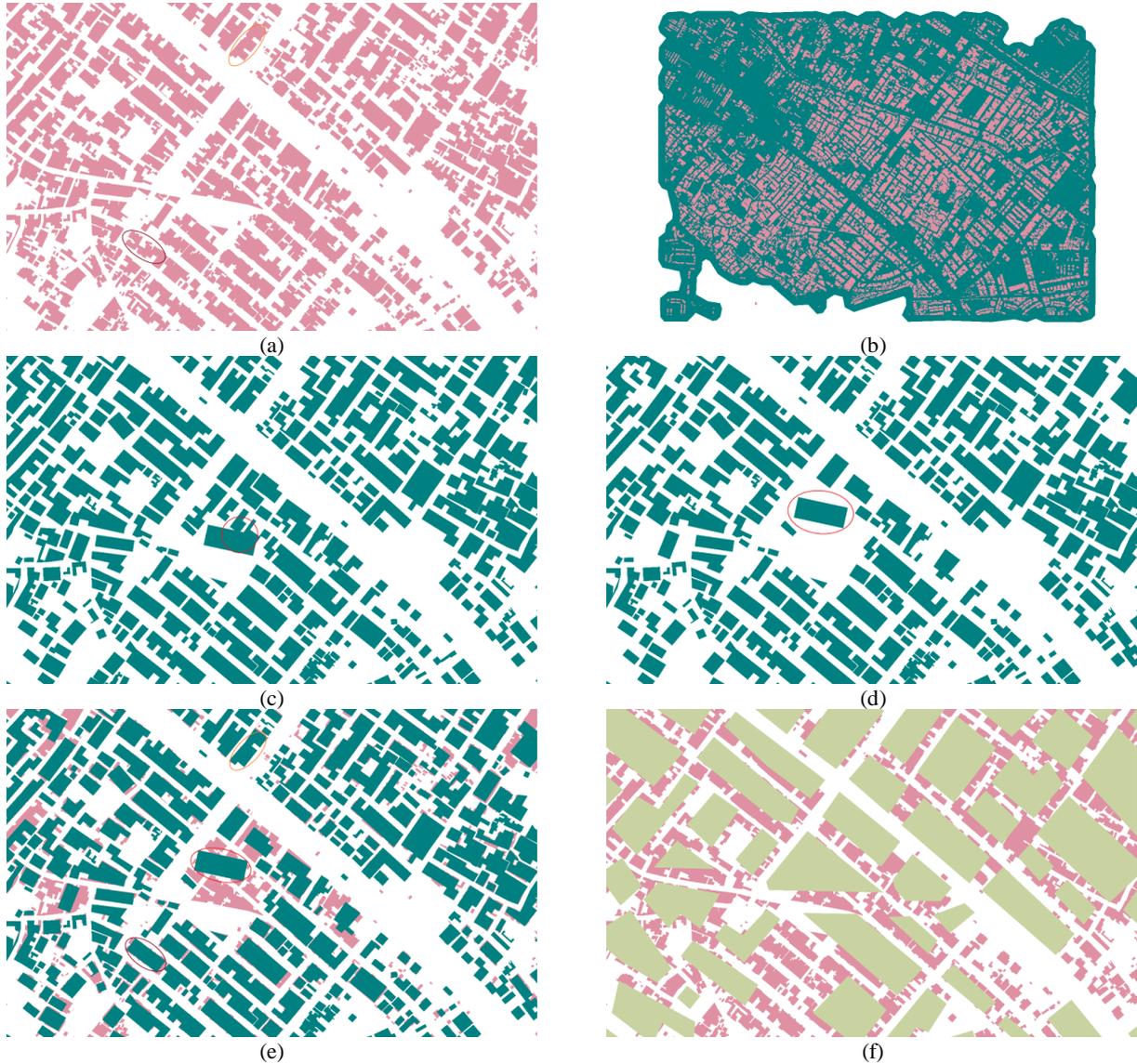

Figure 9. (a) Input SoI 10K data, (b) Block created using enrichment, (c) Output of the AGENT model using building micro agents, (d) Output of the AGENT model using block meso agent, (e) Output of the AGENT model overlaid on SoI 10K, (f) SoI 50K overlaid on SoI 10K.

## 5. CONCLUSIONS AND FUTURE WORK

In this paper, the study has been carried out to test the suitability of CartAGen platform for generalising OS and SoI data products. The merging algorithm is not providing the desired results for OS data products. However, for SoI data product, it is giving improved results compared to that of SoI 50K data. It is identified that the boundaries of buildings, after the merging operation, needs further smoothing. The results of the urban area delimiting algorithm show that the boundaries of the urban areas require simplification and smoothing for better visual clarity at 1:250K representation. From the experiments on the AGENT model, it is realized that the results can be improved by tuning the constraints with appropriate values separately for OS and SoI data products. Furthermore, efforts are being proposed to solve the issues identified in this research as our contribution to the CartAGen platform.

## ACKNOWLEDGMENTS

The authors would like to thank the Ordnance Survey, UK, for their OS OpenData products. We also thank the Survey of India, India for providing the 1:10K scale data. We would like to present our sincerest gratitude to Dr. Guillaume Touya for providing his generous support in the usage of the CartAGen platform.